# The Muon Accelerator Program

Steve Geer
Mail to:  sgeer@fnal.gov
Fermilab, Batavia, IL  60510

Mike Zisman
Mail to:  mszisman@lbl.gov
LBNL, Berkeley, CA XXXX

## 1. Introduction

Multi-TeV Muon Colliders and high intensity Neutrino Factories have captured the imagination of the particle physics community. These new types of facility both require an advanced muon source capable of producing $O(10^{21})$ muons per year. The muons must be captured within bunches, and their phase space manipulated so that they fit within the acceptance of an accelerator. In a Neutrino Factory (NF), muons from this "front-end" are accelerated to a few GeV or a few tens of GeV, and then injected into a storage ring with long straight sections. Muon decays in the straight sections produce an intense neutrino beam. In a Muon Collider (MC) the muons must be cooled by a factor $O(10^6)$ to produce beams that are sufficiently bright to facilitate high luminosities in the collider. Bunches of positive and negative muons are then accelerated to high energy, and injected in opposite directions into a storage ring in which they collide at one or more interaction points.

Over the last decade our understanding of the concepts and technologies needed for Muon Collliders and Neutrino Factories has advanced, and it is now believed that, within a few years, with a well focused R&D effort (i) a Neutrino Factory could be proposed, and (ii) enough could be known about the technologies needed for a Muon Collider to assess the feasibility and cost of this new type of facility, and to make a detailed plan for the remaining R&D.  Although these next NF and MC steps are achievable, they are also ambitious, and will require an efficient organization to accomplish the desired goals with limited resources. The Muon Accelerator Program (MAP) has recently been created to propose and execute this R&D program.

## 2. The Birth of MAP

The Muon Collider Collaboration was formed in 1996, and consisted of about 100 particle physicists and accelerator scientists and engineers from U.S. laboratories and universities. The initial work on the overall Muon Collider concept resulted in the "Muon Collider Feasibility Study Report" in June 1996 [1]. The Neutrino Factory concept emerged in 1998 [2]. The collaboration was subsequently renamed the "Neutrino Factory and Muon Collider Collaboration" (NFMCC). From 1997 to 2010 the NFMCC pursued both NF and MC design and simulation studies [3–5], together with component development and proof-of-principle demonstration experiments. In late 2006, the Muon Collider R&D effort was complemented by the addition of the Muon Collider Task Force (MCTF) hosted by Fermilab. This approximately doubled the support in the U.S. for NF and MC R&D. By 2009 the NFMCC + MCTF community, together with their international partners (MICE [6], EMMA [7], MERIT [8], IDS-NF [9]) had made significant progress, completing a series of NF design feasibility studies [3], completing the proof-of-principle target experiment MERIT, launching the Muon Ionization Cooling Experiment (MICE), initiating a hardware component development program (MuCool), building the MuCool Test Area at Fermilab, and making steady progress with muon cooling channel studies.

Given these achievements, in October 2009 the DOE requested the Fermilab Director to put in place and host a new national Muon Accelerator R&D organization (Muon Accelerator Program, MAP [10]) to replace and streamline the NFMCC+MCTF activities, with an expectation of increased funding. MAP is now in place and functioning. The MAP R&D plan was reviewed in August 2010, and MAP became a formal and fully functional entity with the signing of its management plan in March 2011.

### 3. MAP Mission and Goals

The MAP mission is described in a mission statement:

"…to develop and demonstrate the concepts and critical technologies required to produce, capture, condition, accelerate, and store intense beams of muons for Muon Colliders and Neutrino Factories. The goal of MAP is to deliver results that will permit the high-energy physics community to make an informed choice of the optimal path to a high-energy lepton collider and/or a next-generation neutrino beam facility. Coordination with the parallel Muon Collider Physics and Detector Study and with the International Design Study of a Neutrino Factory will ensure MAP responsiveness to physics requirements."

To accomplish this mission, the main desired MAP R&D deliverables over the next few years are:

i) A Design Feasibility Study Report for a multi-TeV MC including an end-to-end simulation of the MC accelerator complex using demonstrated, or likely soon-to-be-demonstrated, technologies, an indicative cost range, and an identification of further technology R&D that should be pursued to improve the performance and/or the cost effectiveness of the design.
ii) Technology development and system tests that are needed to inform the MC-design feasibility studies, and enable an initial down-selection of candidate technologies for the required ionization cooling and acceleration systems.
iii) Contributions to the International Neutrino Factory Design Study (IDS-NF) to produce a Reference Design Report (RDR) for a NF.

**4.   Achievements in the First Year**

The present fiscal year (FY11) has been a transition year for MAP. It is the first year that the new organization has been responsible for formulating the plan and steering the R&D. In addition to establishing the organization, some significant FY11 milestones have been accomplished:

i) **A new beamline.** The Mucool test Area (MTA) is a critical R&D facility at Fermilab built to enable tests of muon cooling channel components. In FY11, MAP has succeeded in completing and commissioning a beamline from the end of the Fermilab linac to the MTA. This is a major enhancement of the test facility. The first measurements in the beam have also been made (see item 2).

ii) **Cooling channel RF technologies; narrowing the options.** It has been found that the maximum operating gradient for normal conducting copper vacuum cavities is reduced when operated in an axial magnetic field of a few Tesla [11]. If our present designs for muon cooling channels are to work, we must find a way to mitigate this effect. At the end of FY10 there were four different ideas on how to solve the problem. The MAP strategy is to build hardware to test these ideas in the MTA, and to do this as rapidly as is practical, until a solution emerges. In FY11 one of the ideas (so called magnetic insulation) was tested in the MTA. These tests resulted in a deeper understanding of what happens with dark current electrons in the appropriate magnetic geometry, and have enabled the magnetic insulation option to be eliminated from the list of candidate solutions. A second candidate solution is to use a normal conducting cavity filled with hydrogen gas at high pressure. This technology had already been shown to work in a magnetic field, but had not been tested with an ionizing beam. In FY11, the first measurements of a test cavity of this type have been made using the new MTA beam. Further measurements, with both the magnetic field and beam, are expected soon. Finally, preparations are being made for testing next year a third candidate solution, which uses beryllium within the cavity in regions of high surface field. Thus, we anticipate that within the first 18 months, MAP will have tested three out of the four proposed ways to mitigate the effects of magnetic field on RF operation.

iii) **MICE magnets.** The MICE experiment at RAL will test a short section of a muon cooling channel in a muon beam, measuring the response of each incoming muon to the channel. This experiment not only provides an essential test of the simulation programs used to design and study muon cooling channels, but will also provide a technical demonstration of the cooling channel technologies. The large spectrometer solenoids used in the experiment have proven to be more challenging than originally anticipated, and need to be modified. In FY11 significant engineering studies were completed to evaluate the needed modifications, and a plan to complete the modifications established. Execution of the plan should be completed in FY12. This prepares the way for testing a cooling channel section in a muon beam in FY14-15.

iv) **IDS-NF Interim Report.** The NF design studies are being pursued within the framework of the "International Design Study for a Neutrino Factory" which, as its name suggests, is a fully international endeavor, and includes many significant MAP contributions. The IDS-NF desires to produce a "Reference Design Report" in two years time. The milestone this year was to produce an interim report. This report was successfully completed [12] and reviewed by ECFA.

v) **Significant improvements in MC design.** Muon Collider design and simulation studies are essential to understand what hardware is needed to achieve a given performance, to guide R&D priorities, and to understand the implications of measured component performance. The MC design and simulation studies are naturally focused on the things that are unique to muon colliders. In FY11 these studies [13] resulted in (i) an updated design for the proton target area [14] to enable the target area solenoids to work in the hostile radiation environment created by the interaction of a 4MW primary beam, (ii) a first design of a system to separate the positive from the negative muons before they are cooled, which is needed for most cooling channel designs, (iii) improved designs for merging within the cooling channel a string of muon bunches into a single bunch, needed to increase collider luminosity, (iv) improved collider lattice designs that include magnet studies to understand options that enable the dipoles and quadrupoles to operate in an environment in which the muons are decaying to produce high energy electrons, and (v) significant advances in machine-detector interface studies (see 7 below).

vi) **Establishing a vision for upgrading Project X.** One of the motivations for a new high power proton source at Fermilab is to facilitate options for the laboratory's long-term future, and specifically for a Neutrino Factories and/or Muon Collider. It is therefore important that there is a clear concept of how Project X [15] might evolve into a 4MW proton source with the bunch structure required by a NF and/or MC. To arrive at this concept requires a joint effort between the Project X and MAP teams. To facilitate this a MAP-Project X task force was formed, and has already developed an initial concept for the required Project X upgrade.

vii) **Reaching out to the community: machine-detector interface studies and the Muon Collider 2011 meeting.** As the MC accelerator R&D proceeds it is important that there is a good connection with the particle physics community. To facilitate this (i) within MAP there is a machine-detector interface group that iterates on the final focus design and provides background files for detector studies, (ii) a new MC physics and detector group has been begun outside of MAP to engage theorists and detector experts in assessing MC detector performance and final focus requirements, and (iii) a community-wide MC meeting was held in June 2011 (Muon Collider 2011 [16]) to bring together people interested in the accelerator R&D, the detectors, and the physics program, and discuss status and R&D opportunities.

## 5. A Vision for the Future

MAP is focused on developing muon-based options for the future of particle physics. In the U.S., this is coupled to a vision for the future of the accelerator complex at Fermilab. That vision is illustrated in Figures 1-3. It begins with Project X, a new proton source that can support a world class intensity frontier physics program (Fig. 1). The next possible step would be to upgrade Project X to provide a 4MW beam, and add a NF (Fig. 2). Finally, the complex could be upgraded to add a site-filling MC with a center-of-mass energy of 3-4 TeV (Fig. 3).

## 6. Summary

MAP is a new R&D organization that has been established to pursue Muon Collider and Neutrino Factory R&D. The planned MAP R&D builds upon progress over the last decade. In its first year of existence MAP has integrated the ongoing R&D from the NFMCC and MCTF into one coherent effort, and has made some significant progress. The R&D program is ambitious, but is motivated by a strong muon-based vision for the future.

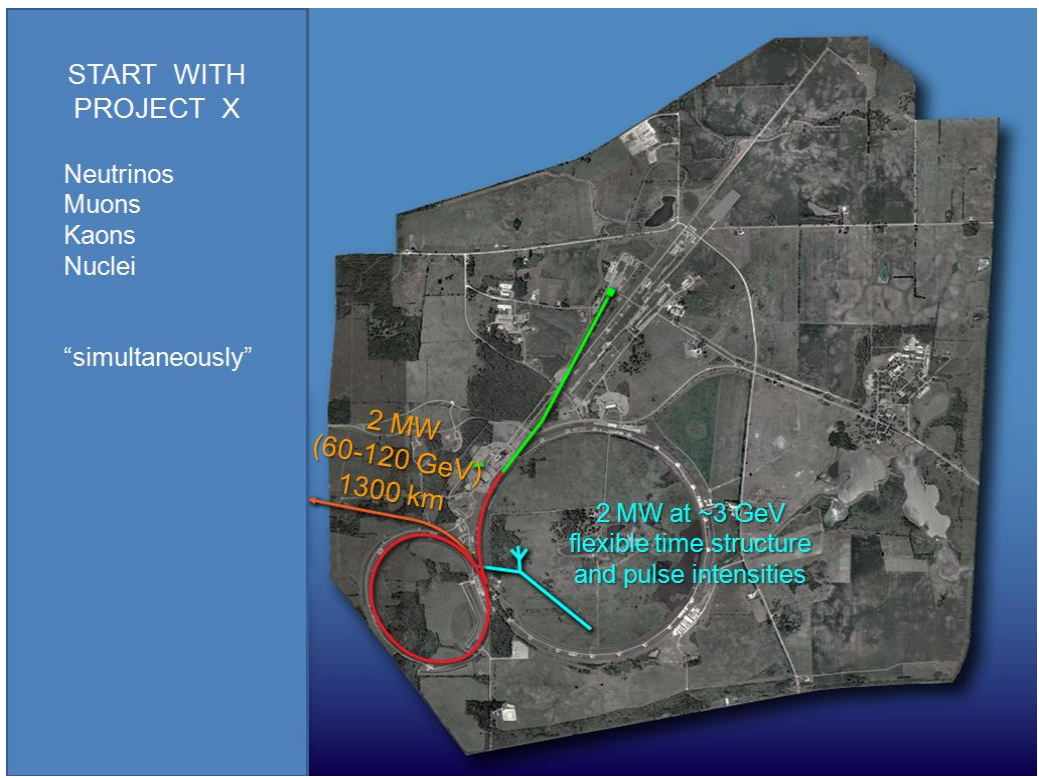

**Figure 1**: A vision for the future: Project X (Step 1).

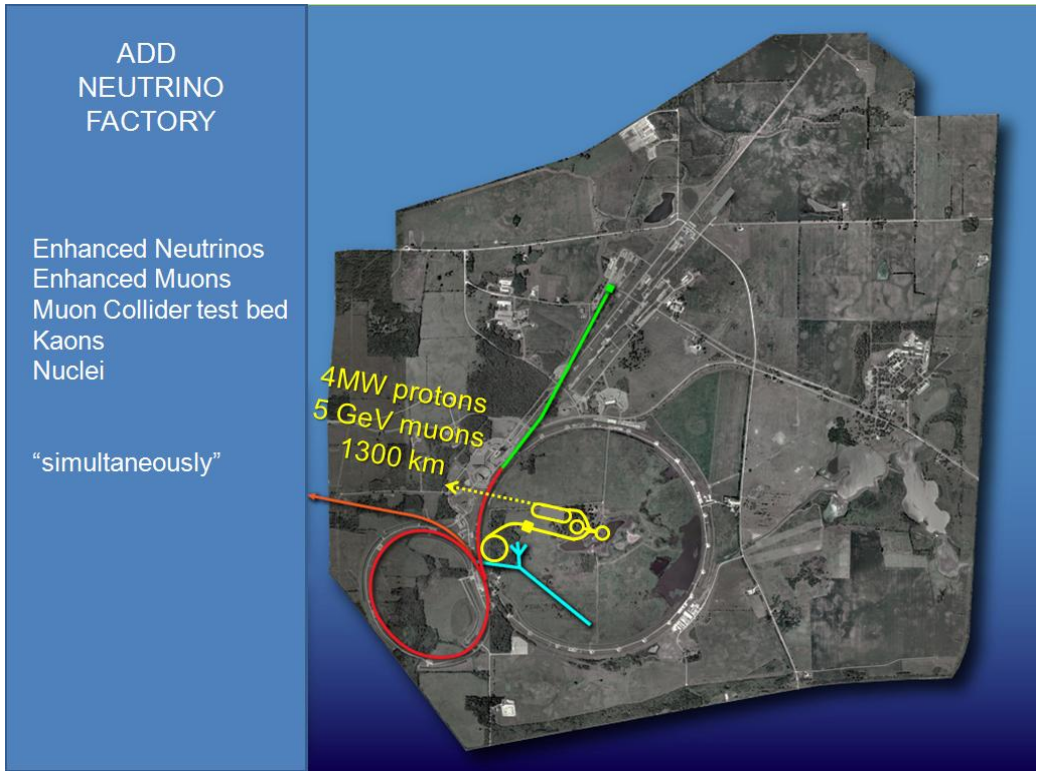

**Figure 2**: A vision for the future: A Neutrino Factory (Step 2).

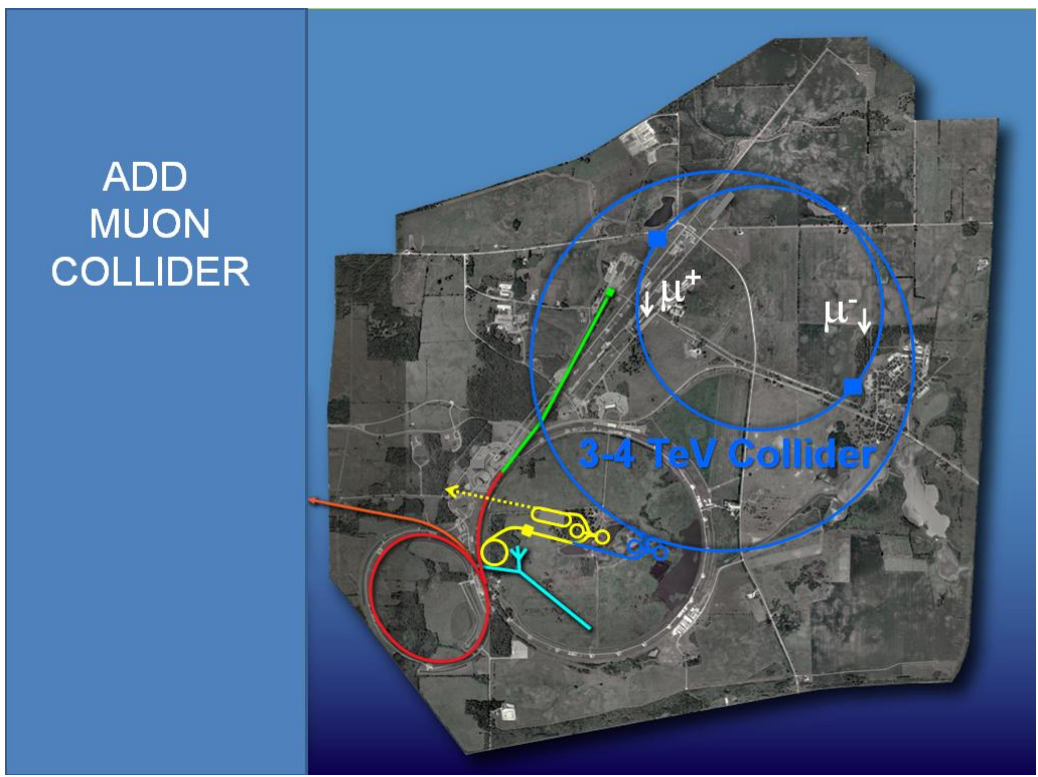

**Figure 3**: A vision for the future: A multi-TeV Muon Collider (Step 3).